\documentclass[11pt,preprint]{article}
\pdfoutput=1
\usepackage{jheppub-arxiv}
\usepackage{graphicx}
\usepackage{tocloft}
\usepackage{appendix}
\usepackage{tikz}
\usepackage{subcaption}
\usepackage{xcolor}
\usepackage{cleveref}
\usetikzlibrary{snakes, patterns,angles,quotes,arrows.meta,shapes.misc}

\title{Bootstrapping Gravity with Crossing Symmetric Dispersion Relations}
\author{Celina Pasiecznik}
\affiliation{Department of Physics, McGill University, 3600 Rue University, Montr\'eal, H3A 2T8, QC Canada
}
\emailAdd{celina.pasiecznik@mail.mcgill.ca}
\date{ }

\abstract{
We derive bounds on Wilson coefficients in gravitational effective field theories using fully crossing symmetric dispersion relations. These sum rules naturally isolate finite subsets of low-energy couplings without relying on the forward limit or specific high-energy completions. We validate our method by matching bounds computed previously for scalar scattering with gravity as well as for graviton scattering. For graviton scattering we construct crossing symmetric combinations of the maximal helicity violating amplitude. We also derive new bounds on the coupling of gravitons to a massive spin-4 state at tree level. 
These results demonstrate the power of crossing symmetric sum rules as a tool in the S-matrix bootstrap.}

\begin{document}

\maketitle
\newpage
\tableofcontents

\section{Introduction}

Effective field theories (EFTs) describe a physical theory at a chosen energy scale without precise knowledge of the physics at very short distances. 
EFTs parametrize low-energy physics via the use of local higher-dimensional operators, each characterized by a Wilson coefficient whose value depends on the precise ultraviolet (UV) completion.
The S-matrix bootstrap program for gravitational EFTs uses fundamental principles, namely analyticity, unitarity, and crossing symmetry, to constrain these Wilson coefficients. 
In practice, the dual approach which makes use of positivity and dispersion relations is able to rule out values of Wilson coefficients of UV theories that do not obey these principles \cite{Adams:2006sv, deRham:2017avq,deRham:2018qqo, Bellazzini:2019xts, Herrero-Valea:2022lfd, Komargodski:2011vj,Luty:2012ww,Tokuda:2020mlf,Alberte:2020jsk,Pajer:2020wnj,Grall:2021xxm,Manohar:2008tc,Low:2009di,Bellazzini:2016xrt,Cheung:2016yqr,Bellazzini:2017fep,Chiang:2022ltp, Alberte:2020bdz,Arkani-Hamed:2020blm, Bellazzini:2019bzh,Berman:2024kdh,Berman:2024eid,Berman:2024owc}.
Alternatively, there is the primal bootstrap framework which constructs an explicit analytic and crossing-symmetric ansatz for the amplitude, typically used for non-perturbative approaches \cite{Paulos:2017fhb, Guerrieri:2020bto,EliasMiro:2022xaa,Paulos:2016but,PhysRevLett.122.241604,Hebbar:2020ukp,Karateev:2022jdb,Chen:2022nym,Cordova:2023wjp,Cordova:2025bah,Correia:2025uvc,Bern:2022yes}.

For gravitational EFTs, these bootstrap constraints are applied to low-energy theories whose leading interaction is Einstein gravity, supplemented by higher-derivative corrections.
At large distances, gravity is described by Einstein's theory of general relativity, which must be UV completed at short distances. 
The space of allowed gravitational UV completions has previously been studied for scalar scattering \cite{sharpboundaries}, for supergravity theories \cite{Albert:2024yap}, and for graviton scattering \cite{Caron_Huot_20232, Caron-Huot:2022jli, Chowdhury:2021ynh}. 
Appropriate assumptions are used for the high-energy behavior of the scattering amplitudes to write dispersion relations that relate the couplings appearing in the low-energy EFT to the high-energy side (see e.g. \cite{Haring:2022cyf}). 
In theories without gravity, the forward limit has been used to isolate a finite subset of couplings to bound on the low-energy side for massless scalars \cite{Bellazzini:2020cot, Tolley_2021, Caron_Huot_2021} and photons \cite{Henriksson:2021ymi, Haring:2022sdp, Henriksson:2022oeu}. 
Crossing symmetry, which follows from relativistic causality, has been used to impose null constraints that strengthen bounds on the couplings \cite{Caron_Huot_2021,Chiang:2021ziz, Fernandez:2022kzi, Berman:2023jys, Albert:2023jtd, Chiang:2023quf}. 
However, in theories with gravity, the graviton pole obstructs the forward limit and \emph{smearing} must be applied to the functionals. 
Carefully constructed (non-trivial) combinations of sum rules, referred to as improved sum rules \cite{sharpboundaries}, only use the forward limit for higher-dimensional contact terms and are designed to isolate the desired couplings. 

An alternative to using improved sum rules is to impose crossing symmetry at the level of the dispersion relation.
In this paper, we show that a more direct way to isolate a finite set of couplings is to use fully crossing symmetric dispersion relations that depend on a variable $z$ which is related to the Mandelstam invariants as given in \cite{Sinha_2021}. 
As we will see, these sum rules \emph{naturally} provide us with null constraints. 
Crossing symmetric sum rules have been used to derive bounds in \cite{Chowdhury:2021ynh,deRham:2022gfe,Li:2023qzs, Beadle:2025cdx, Chang:2025cxc, Bhat:2025zex}.
In contrast to \cite{ Beadle:2025cdx, Chang:2025cxc}, we show that we can exactly reproduce bounds at tree level in flat space for scalar scattering with gravity that match the bounds found using improved sum rules in \cite{sharpboundaries}. 
We compute bounds for graviton scattering in supergravity that match those found in \cite{Albert:2024yap}. 
We also show how to write crossing symmetric sum rules for graviton scattering amplitudes with various helicity configurations. 
Using these sum rules, we compute bounds for graviton scattering that match those found in \cite{Caron_Huot_20232}. 
Crossing symmetric dispersion relations also provide a solution to computing bounds at loop level for any EFT where light states running in the loop lead to singularities in the forward limit,  even without gravity present. 
Thus, the crossing symmetric approach also offers this advantage over improved sum rules which are not stable for loops.

In addition to bounding Wilson coefficients of local contact terms, these crossing symmetric sum rules can also incorporate assumptions about the spectrum of massive states.
In particular, we can include the explicit exchange of a massive higher-spin state and bound its coupling to the external scattered particles.
For example, for QCD at large $N$, bounds were derived in \cite{Albert:2022oes} on the coupling of pions to the rho meson as a function of the ratio of the meson's mass to the EFT cutoff $M$.
Furthermore, massive higher-spin exchanges were included at tree level in \cite{Berman:2024wyt} to study couplings in the effective theory of supersymmetric Yang-Mills.
Mass spectra have also been supplied as inputs in the primal approach in \cite{Cheung:2023adk, Haring:2023zwu}.
We show in the case of supergravity that the crossing symmetric sum rules reproduce the bounds found in \cite{Albert:2024yap} on the coupling of external gravitons to a heavy spin-4 state exchanged in $D=10$. 
For supergravity, we use Gegenbauer polynomials for the partial wave representation of the exchanged massive state. 
Inspired by the supergravity case, we integrate in the next higher-spin massive particle in the $2\rightarrow 2$ effective theory of gravity for the maximal helicity violating (MHV) amplitude  in $D=4$. We use Wigner-d functions for the partial wave representation of the heavy state to account for the helicity of the external gravitons.
We derive bounds on the coupling of external gravitons to the heavy spin-4 state as a function of the ratio of the spin-4 mass to the cutoff $M$, where $M$ corresponds to the mass of the next higher-spin exchanged state. 
Together, these examples demonstrate that crossing symmetric sum rules provide an effective tool in the S-matrix bootstrap.

The paper is organized as follows. Section \ref{sec:setup} reviews the assumptions and setup used for the gravitational EFT bootstrap as well as introduces crossing symmetric dispersion relations. Sections \ref{sec:boundswgrav} and \ref{sec:boundsspectral} contain results that have been computed using the crossing symmetric sum rules which match the literature. 
Section \ref{sec:gravbounds} constructs crossing symmetric combinations of MHV graviton scattering amplitudes, demonstrating how the formalism can be applied to spinning external states.
We first reproduce known bounds on the gravitational Wilson coefficients and then use the same framework to derive new bounds on the coupling of a spin-4 state to external gravitons.
We end with a brief discussion in section \ref{sec:discussion}. Appendix \ref{sec:numerics} provides details on the numerical implementation.

\section{Setup}
\label{sec:setup}
In this section, we state our assumptions and provide the setup for bounding couplings of EFTs that include gravitational interactions.
We are interested in bounding couplings at tree level in $2\rightarrow2$ scattering of scalars or gravitons.
We will use real massless scalar scattering as the simplest example to illustrate our setup.

\subsection{Assumptions}
Our objective is to bound ratios of couplings without assuming a particular high-energy theory, relying only on the assumption that a causal and unitary high-energy completion exists. 
For $2\rightarrow2$ particle scattering, using the all incoming convention, the Mandelstam invariants are
\begin{equation}
s=-\left(p_1+p_2\right)^2, \quad t=-\left(p_2+p_3\right)^2, \quad u=-\left(p_1+p_3\right)^2
\end{equation}
which, for massless particles, satisfy $s+t+u=0$.
We assume the EFT is weakly coupled and thus work at tree level, ignoring any low-energy loop effects. This means the low-energy amplitude only contains terms corresponding to the explicit exchange of particles as well as terms from higher-dimension operators. 
For example, for massless scalar scattering where the scalar is coupled to gravity, the low-energy amplitude is
\begin{equation}
\label{eq:amplow_scalargrav}
\begin{aligned}
\mathcal{M}_{\text {low}}(s, u)= & 8 \pi G\left(\frac{s t}{u}+\frac{s u}{t}+\frac{t u}{s}\right)-g^2\left(\frac{1}{s}+\frac{1}{t}+\frac{1}{u}\right)\\
& -g_0+g_2\left(s^2+t^2+u^2\right)+g_3(s t u)+g_4\left(s^2+t^2+u^2\right)^2+\ldots\,.
\end{aligned}
\end{equation}
In the physical region where $s>0$ and $-s<u<0$, the center-of-mass energy squared is $s$, the squared momentum transfer is $-u$, and the scattering angle is
\begin{equation}
\cos \theta=1+\frac{2 u}{s}\,.
\end{equation}
At high energies, we use a partial wave decomposition of the amplitude. For scalar scattering we use\,\cite{sharpboundaries}
\begin{equation}
\label{eq:partialwaveequation}
\mathcal{M}(s, u)=s^{\frac{4-D}{2}} \sum_{J} n_J^{(D)} c_J(s) \hspace{2pt}\mathcal{P}_J\!\left(1+\frac{2 u}{s}\right)
\end{equation}
where $D$ is the spacetime dimension, $n_J$ is a normalization factor for the partial waves $\mathcal{P}_J$, and $J$ is the spin of the exchanged particle. When the external states are gravitons rather than scalars, we use  Wigner-d functions given in sec.~\ref{sec:gravitonsetup}.
The normalization factor ensures unitarity of the S-matrix and is defined as \cite{Correia:2020xtr}
\begin{equation}
n_J^{(D)} \equiv \frac{2^D \pi^{\frac{D-2}{2}}}{\Gamma\left(\frac{D-2}{2}\right)}(J+1)_{D-4}(2 J+D-3)\,.
\end{equation}
For massless scalar scattering, the partial waves are Gegenbauer polynomials:
\begin{equation}
\label{gegenbauerdefinition}
\mathcal{P}_J(x) \equiv{ }_2 F_1\left(-J, J+D-3, \frac{D-2}{2}, \frac{1-x}{2}\right) \,.
\end{equation}
The spectral density for scalar scattering is defined as $\rho_J(s) = \operatorname{Im} c_J(s)$, which implies the imaginary part of the amplitude can be expressed as
\begin{equation}
\label{eq:partialwaveequationIM}
\operatorname{Im} \mathcal{M}(s, u)=s^{\frac{4-D}{2}} \sum_{J \text { even }} n_J^{(D)} \rho_J(s) \mathcal{P}_J\!\left(1+\frac{2 u}{s}\right)\,.
\end{equation}
Unitarity implies that for all positive energies $s$ and even spin $J$, we have
\begin{equation}
\label{eq:posspectraldensity}
    0 \leq \rho_J(s) \leq 2 
\end{equation} 
which provides us with crucial positivity constraints for the high-energy side of the theory.

In gravitational theories we do \emph{not} assume a Froissart-Martin-like bound.\footnote{The Froissart-Martin bound, $\mathcal{M}(s, t) \lesssim s \log ^{D-2} s$, has been proven for theories with a mass gap \cite{PhysRev.123.1053, PhysRev.129.1432} but has not been proven more generally for theories with massless particles (see \cite{Haring:2022cyf} for $D > 4$ in the case of gravity).} Instead we follow the method of \cite{sharpboundaries, Caron_Huot_20232} and use functionals $f(b)$ that are localized at small impact parameter. 
For fixed impact parameter $b=2J/\sqrt{s}$, the Gegenbauer functions in the large $\sqrt{s}$ limit become Bessel functions \cite{sharpboundaries}
\begin{equation}
\label{besselfunc}
\lim _{m \rightarrow \infty} \mathcal{P}_{\frac{m b}{2}}\left(1-\frac{2 p^2}{m^2}\right)=\frac{\Gamma\left(\frac{D-2}{2}\right)}{(b p / 2)^{\frac{D-4}{2}}} J_{\frac{D-4}{2}}(b p) \,.
\end{equation}
Recall that impact parameter space is Fourier conjugate to momentum space. In momentum space, we define smeared amplitudes as in \cite{sharpboundaries}:
\begin{equation}
\label{eq:smearedamplitude}
\mathcal{M}_{f}(s)=\int_0^{p_\text{max}} d p f(p) \mathcal{M}\left(s,-p^2\right) .
\end{equation}
We suppress the graviton pole, which diverges in the forward limit, by integrating the amplitude against functionals that are localized in momentum space and decay rapidly at large impact parameters, $b$. Physically, this approach allows us to measure EFT couplings at small impact parameters, $b \sim 1/M$, ensuring that the sum rules hold at the scale $M$.
We choose a basis of functions of momentum $p$,
\begin{equation}
\label{smearfunc}
f(p)=\sum_{n=n_\text{min}}^{n_\text{max}} a_n (1-p^2) p^n \,,
\end{equation}
where $n_\text{min}$ and $n_\text{max}$ are chosen for each sum rule.
For $s>M^2$, we apply the partial wave expansion to the smeared amplitudes (\ref{eq:smearedamplitude}). The rapid decay in impact parameter space at some $b_*$ of the functionals $f(p)$ ensures that the sum in (\ref{eq:partialwaveequation}) is restricted to $J \leq \sqrt{s} b_*$ which gives \cite{Caron_Huot_20232}
\begin{equation}
\left|\mathcal{M}_{f}(s)\right| \leq|s| \times \text { constant }
\end{equation}
for real $s>M^2$.
By assuming there are no causality violations for complex energies, we can extend this bound to the entire complex plane, 
\begin{equation}
\label{eq:ampbound}
\lim _{|s| \rightarrow \infty} \frac{\mathcal{M}_{f}(s,u)}{|s|^2} = 0 \text{\quad for fixed $u<0$.}
\end{equation}
We can now use the convergence and analyticity of the amplitude to relate low and high energies through dispersion relations.
We will use fully crossing symmetric dispersion relations introduced in sec.\,\ref{sec:crosssetup} which \emph{naturally} select a finite number of couplings on the low-energy side without making use of the forward limit.

\subsection{Crossing symmetric dispersion relations}
\label{sec:crosssetup}
We use crossing symmetric dispersion relations from  \cite{Sinha_2021}, originally defined in \cite{PhysRevD.6.2953,MAHOUX1974297}. These dispersion relations are written in terms of a complex variable $z$ and an auxiliary momentum variable $p$. The relation between the Mandelstam variables and $(z,p)$ is
\begin{equation}
\label{stucrossvar}
s_n=-p^2+\frac{p^2\left(z-z_n\right)^3}{z^3-1}, \quad n=1,2,3, \quad\text{and}\quad
\frac{1}{p^2}= -\frac{1}{s_1}-\frac{1}{s_2}-\frac{1}{s_3}
\end{equation}
where $z_n$ are cubic roots of unity, namely $z_1 =1, z_2 = e^{2 \pi i/3}, z_3 =e^{4 \pi i/3}$. Here $s_1 = s - \frac{4 \mu}{3}$, $s_2 = t- \frac{4 \mu}{3}$, $s_3 = u- \frac{4 \mu}{3}$, where we take $\mu=0$ since we will only be considering massless particles for the external states.
The Regge limit in the three channels corresponds to the points $z_n$ on the unit circle as depicted in fig.~\ref{fig:contour_comparison}, around which the integration contour is taken. 
The massless poles appearing in the low-energy amplitude correspond to $z=0$ and $z=\infty$. 
We can write crossing symmetric sum rules using these new variables for fixed-$p$ as
\begin{equation}
\label{eq:sumrules}
B_k(p^2) \equiv \oint_{z=z_1,z_2,z_3} \frac{d z}{4 \pi i}  \mathcal{K}_k(z,p^2) \mathcal{M}(z, p^2)=0, 
\end{equation}
where the amplitude $\mathcal{M}(z, p^2)$ must be fully crossing symmetric. The crossing symmetric kernel is
\begin{equation}
\label{eq:crossingkernel}
\mathcal{K}_k(z,p^2)=\frac{(1 + z^3)}{z(1 - z^3)} \left( \frac{(1 - z^3)^2}{- 27 p^4  z^3} \right)^{\frac{k}{2}}\,, \quad\quad k=2,4 \ldots\,
\end{equation}
where $k$ labels the subtraction order, or equivalently which order in the EFT expansion is being probed.
The high-energy branch cuts lie in the complex $z$ plane as shown in fig.\,\ref{fig:contour_comparison}. 
\begin{figure}
    \centering
    \begin{minipage}{0.45\textwidth}
        \centering
        \scalebox{1.2}{\begin{tikzpicture}
\draw (0, 0) circle [radius=2cm];
    \draw[red, decorate, decoration={snake, amplitude=0.5mm, segment length=2mm}] 
        ({1.99*cos(0)},{1.99*sin(0)}) 
        arc (0:122.3:1.98);
    \draw[red, decorate, decoration={snake, amplitude=0.5mm, segment length=2mm}] 
        ({1.99*cos(242)},{1.99*sin(242)}) 
        arc (242:360:1.97);
    \draw[red, decorate, decoration={snake, amplitude=0.5mm, segment length=2mm}] 
    ({1.99*cos(120)},{1.99*sin(120)}) 
        arc (120:242.5:1.98);
        
\draw[fill=white] (-1, 1.73) circle [radius=0.1cm];
\draw[fill=white] (2, 0) circle [radius=0.1cm]; 
\draw[fill=white] (-1, -1.73) circle [radius=0.1cm]; 

        \fill (-1, -1.73) circle [radius=0.03cm] node[below left] {\scriptsize fixed-$t$};
        \fill (2, 0) circle [radius=0.03cm] node[right] {\scriptsize fixed-$s$};
        \fill (-1, 1.73) circle [radius=0.03cm] node[above left] {\scriptsize fixed-$u$};

    \draw[] (2.2, 2.5) -- (2.2, 2.2) -- (2.5, 2.2);
    \node at (2.35, 2.35) {\textit{z}};

\node[draw=black, cross out, inner sep=1pt, thick,label=below:{\scriptsize $\infty$}] at (2.35,1.5) {};
\node[draw=black, cross out, inner sep=1pt, thick,label=below:{\scriptsize $0$}] at (0,0) {};
    
\end{tikzpicture}}
        \label{fig:contourzplane2} 
    \end{minipage}
    \hfill
    \begin{minipage}{0.45\textwidth}
        \centering
        \scalebox{1.2}{\begin{tikzpicture}
\draw (0, 0) circle [radius=2cm];
    \draw[red, decorate, decoration={snake, amplitude=0.5mm, segment length=2mm}] 
        ({1.99*cos(0)},{1.99*sin(0)}) 
        arc (0:122.3:1.98);
    \draw[red, decorate, decoration={snake, amplitude=0.5mm, segment length=2mm}] 
        ({1.99*cos(242)},{1.99*sin(242)}) 
        arc (242:360:1.97);
    \draw[red, decorate, decoration={snake, amplitude=0.5mm, segment length=2mm}] 
    ({1.99*cos(120)},{1.99*sin(120)}) 
        arc (120:242.5:1.98);
        
\draw[fill=white] (-1, 1.73) circle [radius=0.1cm];
\draw[fill=white] (2, 0) circle [radius=0.1cm]; 
\draw[fill=white] (-1, -1.73) circle [radius=0.1cm]; 

        \fill (-1, -1.73) circle [radius=0.03cm] node[below left] {\scriptsize fixed-$t$};
        \fill (2, 0) circle [radius=0.03cm] node[right] {\scriptsize fixed-$s$};
        \fill (-1, 1.73) circle [radius=0.03cm] node[above left] {\scriptsize fixed-$u$};

    \draw[] (2.2, 2.5) -- (2.2, 2.2) -- (2.5, 2.2);
    \node at (2.35, 2.35) {\textit{z}};

\node[draw=black, cross out, inner sep=1pt, thick,label=below:{\scriptsize $\infty$}] at (2.35,1.5) {};
    
\foreach \angle in {60, 180, 300} {
        \draw[red, decorate, decoration={snake, amplitude=0.5mm, segment length=2mm}] ({0.5*cos(\angle)},{0.5*sin(\angle)}) -- ({2.5*cos(\angle)},{2.5*sin(\angle)});
       }

    \node[draw=black, cross out, inner sep=1pt, thick,label=below:{\scriptsize $0$}] at (0,0) {};
\end{tikzpicture}}
        \label{fig:contourzplane3}  
    \end{minipage}
    
    \caption{On the left, $p_\text{max}=M/\sqrt{3}$ leads to the UV branch cuts (in red) for each channel existing entirely on the unit circle. On the right, $p_\text{max}=M$  causes the branch cuts (in red) to stretch onto cuts at $\text{Arg } z=\pi/3, \pi, 5\pi/3$. Note these branch cuts never reach the pole at $z=0$.}
    \label{fig:contour_comparison}
\end{figure}
By deforming the contour \eqref{eq:sumrules}
around the high-energy branch cuts and the low-energy poles, we are able to write dispersion relations:
\begin{equation}
\mathcal{C}_k(p^2) \equiv \oint_{z=0} \frac{d z}{4 \pi i} \mathcal{K}_k(z,p^2) \mathcal{M}\left(z, p^2\right)=-\oint_{C_{\mathrm{UV}}} \frac{d z}{4 \pi i} \mathcal{K}_k(z,p^2) \mathcal{M}\left(z, p^2\right)\,.
\end{equation}
We would like to keep the partial wave decomposition in terms of the Mandelstam invariant $s$ and the auxiliary momentum variable $p$. Thus, we need to write the crossing symmetric kernel $\mathcal{K}_k(z,p^2)$ in $s$, which is 
\begin{equation}
  \mathcal{K}_k(s,p^2)=\frac{2s+3 p^2}{s(s+p^2)}
  \left(\frac{s+p^2}{s^3}\right)^{\frac{k}{2}} \quad\quad k=2,4 \ldots\,. 
\end{equation}
The rest of the Mandelstam invariants  are
\begin{equation}
\label{eq:mandelstamCS}
t=\frac{1}{2} s \left(-1 - \frac{\sqrt{s-3 p^2 }}{\sqrt{s+p^2}}\right), \qquad u= \frac{1}{2} s \left(-1 + \frac{\sqrt{s-3 p^2}}{\sqrt{s+p^2}}\right)\,
\end{equation}
and
\begin{equation}
\label{eq:costhetaCS}
\cos \theta=1+ \frac{2u}{s}=\frac{\sqrt{m^2-3 p^2}}{\sqrt{m^2+p^2}}\,.
\end{equation}
We define heavy averages for the high-energy side as
\begin{equation}
\langle(\cdots)\rangle \equiv \frac{1}{\pi} \sum_{J} n_J^{(D)} \int_{M^2}^{\infty} \frac{d m^2}{m^{2}}\rho_J\left(m^{2}\right)(\cdots)\,.
\end{equation}
The dispersion relations are then
\begin{equation}
\label{eq:crosssym_disprel}
\mathcal{C}_k(p^2) \equiv \underset{z=0}{\text{Res}}\left[\mathcal{K}_k(z,p^2) \mathcal{M}_{\mathrm{low}}\right]
=\left\langle \mathcal{K}_k(m^2,p^2) \hspace{2pt} \pi_J\!\left(\frac{\sqrt{m^2-3 p^2}}{\sqrt{m^2+p^2}}\right)\right\rangle \,
\end{equation}
where $\pi_J$ are generalized partial waves.

Considering scalar scattering with gravity, the crossing symmetric sum rules give
\begin{equation}
\label{eq:scalarsumrules}
\begin{aligned}
    &k=2:  &\frac{8 \pi G}{p^2} +2 g_2 + p^2 g_3&=
    \left\langle
    \frac{2 m^2 + 3 p^2}{m^6} \mathcal{P}_J\left(\frac{\sqrt{m^2-3 p^2}}{\sqrt{m^2+p^2}}\right)\right\rangle\\
    &k=4:  &4 g_4 + 2 p^2 g_5 + p^4 g_6'&= \left\langle
    \frac{(m^2+p^2)(2 m^2 + 3 p^2)}{m^{12}} 
    \mathcal{P}_J\left(\frac{\sqrt{m^2-3 p^2}}{\sqrt{m^2+p^2}}\right)\right\rangle\\
    &k=6:  &8 g_6 + 4 p^2 g_7 + 2 p^4 g_8' + p^6 g_9' &= \left\langle
    \frac{(m^2+p^2)^2(2 m^2 + 3 p^2)}{m^{18}} 
    \mathcal{P}_J\left(\frac{\sqrt{m^2-3 p^2}}{\sqrt{m^2+p^2}}\right)\right\rangle\\
    \end{aligned}
\end{equation}

The advantage of the new sum rules is that for a chosen $k$ in \eqref{eq:crossingkernel}, the low-energy side isolates a finite number of couplings without having to take linear combinations of sum rules. 
Thus, these crossing symmetric sum rules can be used in place of the improved sum rules found in Ref.~\cite{sharpboundaries} and do not rely on taking the forward-limit. 
To reiterate: by making explicit use of fully crossing symmetric dispersion relations, we are able to \emph{naturally} isolate a finite number of couplings on the low-energy side of the sum rules. 

\subsection{Smeared functionals, null constraints, and semidefinite programming}
For the crossing symmetric sum rules, we smear the functionals against wave-packets in \eqref{eq:smearedamplitude} where $p_\text{max}$ is either taken to be $M/\sqrt{3}$ or $M$ as shown in fig.~\ref{fig:contour_comparison} depending on the spectral assumptions used for the high-energy side. The choice of $p_\text{max}$ is explained further in sec.~\ref{sec:spectralassumptions}.

Null constraints are typically derived from crossing symmetry, however, for the crossing symmetric dispersion relations they appear naturally as the higher-$k$ sum rules. Explicitly, we add null constraints by taking linear combinations of functionals at each $k$ which eliminates all unwanted couplings on the low-energy side for that sum rule.
After smearing the dispersive sum rules, we use semidefinite programming, SDPB \cite{simmonsduffin2015semidefinite, Landry:2019qug}, to compute the coupling bounds using the following optimization problem:
\begin{equation}
\begin{aligned}
\text{maximise} \quad & 
\sum_{i}\int_{0}^{M}\!dp\,f_i(p)\,B_{i}(p^{2})
\\[4pt]
\text{subject to} \quad &
\sum_{i}\int_{0}^{M}\!dp\,f_i(p)\,
{B}_{i}\!\bigl[p^{2},m^{2},J,\lambda\bigr]\;\succeq\;0
\quad
\forall\, m>M,\;\rho=[J,\lambda]\,.
\end{aligned}
\end{equation}
The label $i$ enumerates the independent dispersive sum rules.  Intermediate states are organized by little-group representations 
\(\rho=[J,\lambda]\), where
\(\lambda=(m_{2},\dots,m_{n})\)
with \(n=\lfloor(d-1)/2\rfloor\). 
Further details on the mass and spin discretization, the impact-parameter constraints, and the large-$b$ positivity conditions are given in app.~\ref{sec:numerics}.

\section{Bounds with gravity revisited}
\label{sec:boundswgrav}
In this section, we revisit gravitational EFT bounds using crossing symmetric dispersion relations, considering both scalar scattering with gravity and maximal supergravity.

\subsection{Scalar theory without gravity}
As a first example, we take $G=0$ in \eqref{eq:amplow_scalargrav} and evaluate the heavy averages in \eqref{eq:scalarsumrules} on states ($m^2,J$) with
\begin{equation}
    m^2>M^2 \quad \forall J>0, \quad J \text{ even}
\end{equation}
where $M$ is the EFT cutoff.
Taking the forward limit, we get
\begin{equation}
g_2=\left\langle\frac{1}{m^4}\right\rangle, \quad g_3=\left\langle\frac{3-\frac{4}{D-2} \mathcal{J}^2}{m^6}\right\rangle
\end{equation}
which reproduce the exact analytic expressions for the couplings found in \cite{Caron_Huot_2021,sharpboundaries}.

\subsection{Spectral assumptions}
\label{sec:spectralassumptions}
In crossing symmetric variables, the cosine of the scattering angle (\ref{eq:costhetaCS}) will evaluate to imaginary values if we include all states $m> M$ and smear the functionals for momenta values up to the cutoff, $0\leq p\leq M$. 
Evaluating the Gegenbauer polynomials in the example of the scalar theory (\ref{gegenbauerdefinition}) for imaginary arguments produces large oscillating values which SDPB tries to find positive combinations of, leading to large cancellations. 
How large these values are depends on the spin at which the partial wave is evaluated. 
Since the high-energy sum rules should in principle be evaluated over all masses and spins in the spectral density, this can hinder convergence of SDPB. 
To address this issue, we adjust the heavy state list to exclude $M\leq  m\leq \sqrt{3} M$ for $J>J_*$. 
Schematically, this spectral assumption is illustrated by:
\begin{equation}
    \includegraphics[width=0.7\linewidth]{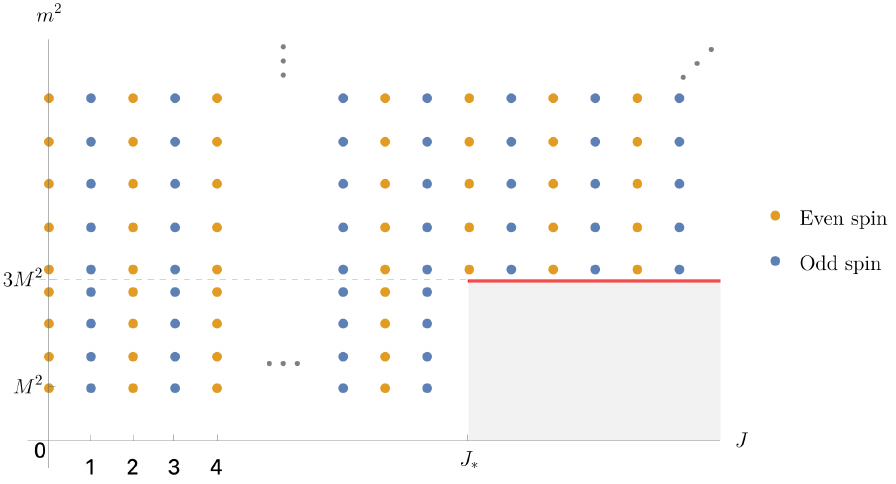}
\end{equation}
Here the heavy state list for $J\leq J_*$ includes all $m\geq M$, and for $J> J_*$, $m\geq \sqrt{3}M$. If we choose $J_*$ to be large enough, we find that such an assumption on the heavy states produces a negligible effect on the coupling bounds. 

\begin{figure}
    \centering
    \includegraphics[width=0.9\linewidth]{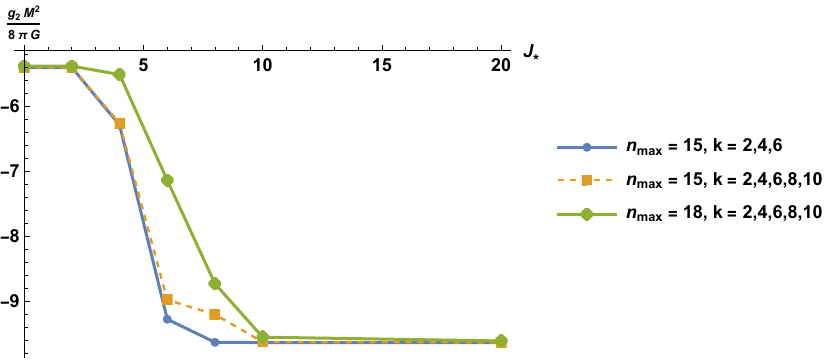}
    \caption{Lower bound on $g_2$ in $D=6$ as a function of the choice of $J_*$ in the spectral assumption. The bounds depend on the choice of $J_*$, as well as on the number of functionals used which is dictated by the number of $k$-sum rules and the choice of $n_\text{max}$ in the smearing (\ref{smearfunc}).}
    \label{fig:Jstarplot}
\end{figure}

As an example, we compute the lower bound for scalar scattering with gravity on $
{g_2 M^2}/{8 \pi G}
$ in $D=6$ as a function of $J_*$ where the heavy states only include even spin. The results are displayed in fig.~\ref{fig:Jstarplot}. We find the bounds converge for a given number of functionals as long as a large enough $J_*$ is used. 

\subsection{Scalar theory with gravity}
For scalar scattering in $D=6$ with $G\neq 0$, we use the spectral assumption described above and smear the functionals over the momentum range $0\leq p\leq p_{\max}$. 
We set $M=1$ and use $J_*=60$. 
The choice of $p_{\max}$ is important. 
To reproduce the bounds obtained from improved sum rules in \cite{sharpboundaries}, one must take $p_{\max}=M$, since in the large-$s$ limit the crossing symmetric kinematics in \eqref{eq:mandelstamCS} gives $u\rightarrow -p^2$.

\begin{figure}[h!]
    \centering
    \includegraphics[width=0.9\linewidth]{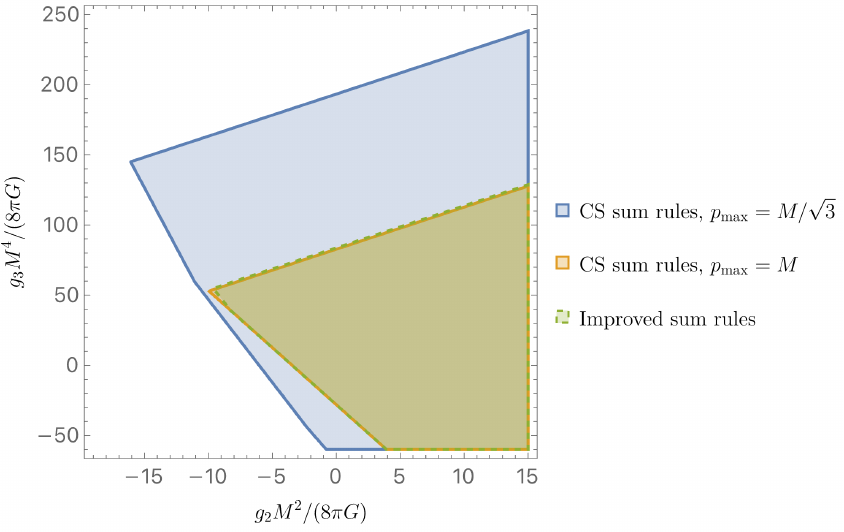}
    \caption{Comparing bounds on $g_2$ vs. $g_3$ in scalar scattering with gravity computed using crossing symmetric dispersion relations with different $p_\text{max}$ and using improved sum rules from \cite{sharpboundaries}.}
    \label{fig:scalargrav}
\end{figure}

In fig.~\ref{fig:scalargrav}, we compare bounds on $g_2$ and $g_3$ obtained using crossing symmetric sum rules with two choices of smearing range. 
Taking $p_{\max}=M$ reproduces the improved-sum-rule bounds of \cite{sharpboundaries}. 
Taking instead $p_{\max}=M/\sqrt{3}$ reproduces the tree-level crossing-symmetric bounds of \cite{Chang:2025cxc, Beadle:2025cdx}. 
The comparison shows that the smaller smearing range gives weaker bounds; smearing up to $p_{\max}=M$ is therefore necessary to recover the optimal bounds obtained from improved sum rules.
For the numerical comparison in fig.~\ref{fig:scalargrav}, we use
\begin{equation}
\begin{aligned}
    \text{CS sum rules:} \qquad &k\in\{2,4,6,8,10\},\qquad n_{\min}=2, \quad n_{\max}=18 \\
    \text{Improved sum rules:} \qquad &k\in\{2,4,6,8,10\},\qquad n_{\min}=2, \quad n_{\max}=15\,.
    \end{aligned}
\end{equation}

\subsection{Maximal supergravity theory}
\label{sec:sugratheory}
We revisit $2\rightarrow2$ scattering of gravitons in $D=10$ maximal supergravity following \cite{sharpboundaries}, for which the graviton exists in the same multiplet as a scalar. The low-energy amplitude is
\begin{equation}
\mathcal{M}_{\text {SUGRA}}^{\text {low}}(s, u)=\frac{8 \pi G}{s t u}+g_0+g_2\left(s^2+t^2+u^2\right)+\ldots\,
\end{equation}
which is crossing symmetric in $s,t,$ and $u$.
In maximal supergravity, supersymmetric Ward identities allow the supergraviton amplitude to be written in terms of this auxiliary fully crossing symmetric scalar amplitude. 
The spin-2 Regge boundedness of the graviton amplitude then translates into the improved behavior
\begin{equation}
\lim_{|s|\rightarrow \infty} 
s^2 \mathcal{M}_{\text{SUGRA}}(s,u)=0,
\qquad \text{for fixed } u<0.
\end{equation} 
Thus, our dispersion relations begin at $k=-2$.
Applying the crossing symmetric sum rules to maximal supergravity scattering gives, for the first few values of $k$,
\begin{equation}
\begin{aligned}
    &k=-2:  &\frac{8 \pi G}{p^2}&=
    \left\langle
    \frac{m^6 \left(2 m^2 + 3 p^2\right)}{\left(m^2 + p^2\right)^2} \mathcal{P}_J\left(\frac{\sqrt{m^2-3 p^2}}{\sqrt{m^2+p^2}}\right)\right\rangle\\
    &k=0:  &g_0 &= \left\langle
    \frac{(2 m^2 + 3 p^2)}{(m^2 + p^2)} 
    \mathcal{P}_J\left(\frac{\sqrt{m^2-3 p^2}}{\sqrt{m^2+p^2}}\right)\right\rangle\\
    &k=2:  &2 g_2 + g_3 p^2 &= \left\langle  \frac{2 m^2 + 3 p^2}{m^6}
    \mathcal{P}_J\left(\frac{\sqrt{m^2-3 p^2}}{\sqrt{m^2+p^2}}\right)\right\rangle\\
    \end{aligned}
\end{equation}
Again we see that only a finite number of couplings appear on the low-energy side of each sum rule.
To bound $g_0$, we use the $k=-2$ relation for the normalization, the $k=0$ sum rule to isolate $g_0$, and the higher-$k$ sum rules as null constraints.
We take
\begin{equation}
\label{eq:sugraparameters}
    k\in\{-2,0,2,4,6\}, \qquad n_{\min}=2, \quad n_{\max}=18\,.
\end{equation}
After smearing these sum rules up to $p_\text{max}=M$, we find
\begin{equation}
\frac{g_0 M^6}{8 \pi G} \leq 2.97 \,.
\end{equation}
The resulting bound matches the result of \cite{Albert:2024yap}. 
The computational simplification here comes from using the fully crossing symmetric setup without applying the forward limit.

\section{Bounds with higher spin spectral assumptions}
\label{sec:boundsspectral}
In this section, we use crossing symmetric sum rules to constrain the coupling between external particles and an explicitly integrated-in massive state.
We first apply this method to maximal supergravity, and later to the exchange of a massive spin-4 state coupled to external gravitons in sec.~\ref{sec:spin4exchange}.

\subsection{Spectral assumptions with integrated-in states}
Following \cite{Albert:2022oes}, we consider an explicitly integrated-in massive state of mass $m_X$ below the new EFT cutoff $M'$, with $M'\geq m_X$.
Using the crossing symmetric variables, the new massive state appears in the complex plane as depicted in fig.~\ref{fig:newzeroscomplexplane} with the branch cuts now beginning at $M'$.
\begin{figure}[h!]
    \centering
    \scalebox{1.3}{\begin{tikzpicture}
\draw (0, 0) circle [radius=2cm];
    \draw[red, decorate, decoration={snake, amplitude=0.5mm, segment length=2mm}] 
        ({1.99*cos(0)},{1.99*sin(0)}) 
        arc (0:122.3:1.98);
    \draw[red, decorate, decoration={snake, amplitude=0.5mm, segment length=2mm}] 
        ({1.99*cos(242)},{1.99*sin(242)}) 
        arc (242:360:1.97);
    \draw[red, decorate, decoration={snake, amplitude=0.5mm, segment length=2mm}] 
    ({1.99*cos(120)},{1.99*sin(120)}) 
        arc (120:242.5:1.98);
        
\draw[fill=white] (-1, 1.73) circle [radius=0.1cm];
\draw[fill=white] (2, 0) circle [radius=0.1cm]; 
\draw[fill=white] (-1, -1.73) circle [radius=0.1cm]; 

        \fill (-1, -1.73) circle [radius=0.03cm] node[below left] {\scriptsize fixed-$t$};
        \fill (2, 0) circle [radius=0.03cm] node[right] {\scriptsize fixed-$s$};
        \fill (-1, 1.73) circle [radius=0.03cm] node[above left] {\scriptsize fixed-$u$};

    \draw[] (2.2, 2.5) -- (2.2, 2.2) -- (2.5, 2.2);
    \node at (2.35, 2.35) {\textit{z}};

\node[draw=black, cross out, inner sep=1pt, thick,label=below:{\scriptsize $\infty$}] at (2.35,1.5) {};
    
\foreach \angle in {60, 180, 300} {
        \draw[red, decorate, decoration={snake, amplitude=0.5mm, segment length=2mm}] ({1*cos(\angle)},{1*sin(\angle)}) -- ({2.5*cos(\angle)},{2.5*sin(\angle)});
     \fill[red] ({0.5*cos(\angle)},{0.5*sin(\angle)}) circle (2pt);
    }

    \node[draw=black, cross out, inner sep=1pt, thick,label=below:{\scriptsize $0$}] at (0,0) {};
\end{tikzpicture}}
    \caption{The red dots denote the poles in the complex $z$-plane corresponding to the $s$-, $t$-, and $u$-channel exchanges of the integrated-in massive state. 
    Three additional poles lie outside the unit circle and are not depicted.}
    \label{fig:newzeroscomplexplane}
\end{figure}
The spectral density used in the partial wave expansion (\ref{eq:partialwaveequationIM}) is adjusted to include a delta function at the mass $m_X$ due to the pole appearing in the low-energy amplitude from the new exchanged state at tree level \cite{albert2023bootstrapping}:
\begin{equation}
\rho_J(s)= g_{X}^2 \frac{\pi m_X^2}{n_J} \delta\left(s-m_X^2\right) \delta_{J, J_X}+\tilde{\rho}_J(s).
\end{equation}
The heavy averages now contain a term corresponding to the spin-$J_X$ state,
\begin{equation}
\left\langle F\left(m^2, J\right)\right\rangle_{m^2 \geq M^2}=g_{X}^2 F\left(m_X^2, J_X\right)+\left\langle F\left(m^2, J\right)\right\rangle_{m^2 \geq {M'}^2}\,.
\end{equation}
This setup is motivated by theories in which part of the massive spectrum is specified explicitly, such as the first massive higher-spin state on a Regge trajectory.
For $M'>m_X$, we assume there are no new states with mass $m_X<m<M'$; hence we assume a mass gap between the integrated-in state and the new cutoff.
We use the crossing symmetric sum rules to bound ratios of couplings involving $g_X^2$, the squared  three-point coupling between the external states and the exchanged massive state.
Since the mass of the integrated-in state is not fixed, we bound $g_X^2$ as a function of the mass ratio $m_X^2/{M'}^2$.

\subsection{Bounds on $g_\phi$ in maximal supergravity revisited}
\label{sec:sugranewstate}
We now return to the $D=10$ maximal supergravity setup of sec.~\ref{sec:sugratheory} and use the crossing symmetric sum rules to bound the coupling of an explicitly integrated-in massive state to the external gravitons.
In the auxiliary scalar amplitude, the first massive state on the leading Regge trajectory appears as an effective spin-0 exchange, corresponding to a physical spin-four state in the supergraviton amplitude. 
We denote its mass by $m_\phi$ and its three-point coupling to the external states by $g_\phi$. 
The low-energy amplitude then contains the explicit pole contribution
\begin{equation}
\mathcal{M}_{\text {SUGRA}}(s, u)=\frac{8 \pi G}{s t u}+g_{\phi}^2\left( \frac{1}{m_{\phi}^2-s} + \frac{1}{m_{\phi}^2-t} + \frac{1}{m_{\phi}^2-u} \right)
+\ldots\,.
\end{equation}

\begin{figure}
    \centering
    \includegraphics[width=0.7\linewidth]{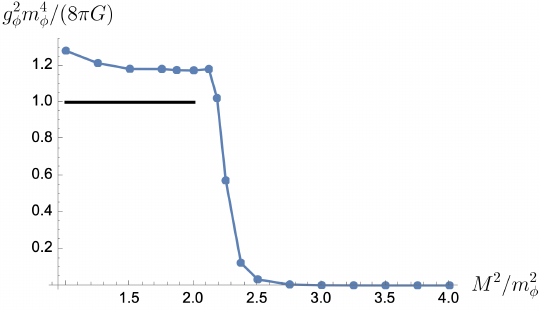}
    \caption{Bound on the coupling of gravitons with the effective spin-0 heavy state in $D=10$ supergravity theory. The horizontal line represents the coupling predicted by the Virasoro-Shapiro amplitude.}
    \label{fig:sugramphi}
\end{figure}

We use crossing symmetric sum rules to bound the coupling $g_\phi^2$ relative to the gravitational coupling $G$. 
Following \cite{Albert:2024yap}, we smear the functionals up to $p_{\max}=m_\phi$ and compute the bound as a function of the ratio $M^2/m_\phi^2$, where $M$ is the cutoff associated with the next higher-spin massive state. 
The numerical setup uses the same values of $k$, $n_{\min}$, and $n_{\max}$ as in (\ref{eq:sugraparameters}).
The resulting bounds are shown in fig.~\ref{fig:sugramphi}; they reproduce the bounds found in \cite{Albert:2024yap}. 
A notable feature is the sharp drop in the bound after $M^2/m_\phi^2\simeq 2$, indicating that a nonzero coupling to the integrated-in state is only compatible with the assumed gap up to this point. 
This is consistent with the interpretation in \cite{Albert:2024yap}, where the extremal solution requires additional states near $m^2\simeq 2m_\phi^2$.
The agreement demonstrates that crossing symmetric dispersion relations can incorporate explicit spectral information, allowing us to bound couplings to integrated-in massive states in gravitational theories.

\section{Bounds for graviton scattering}
\label{sec:gravbounds}
We now apply the crossing symmetric formalism to graviton scattering in $D=4$. 
The main goal of this section is twofold: first, to construct crossing symmetric combinations of MHV graviton amplitudes and reproduce known bounds on higher-derivative corrections to Einstein gravity; second, to integrate in a massive spin-4 state and bound its coupling to external gravitons.

\subsection{Setup}
\label{sec:gravitonsetup}
We consider $2\rightarrow2$ scattering of massless gravitons in $D=4$, following the setup of \cite{Caron_Huot_20232}.  
The low-energy theory consists of Einstein gravity supplemented by higher-derivative corrections, schematically written as
\begin{equation}
S=\frac{1}{16 \pi G} \int d^4 x \sqrt{g}\left(R+g_{R^{(3)}} \operatorname{Riem}^3+g_{R^{(4)}} \operatorname{Riem}^4+\ldots\right)+S_{\text {matter }}
\end{equation}
where $S_{\text {matter }}$ includes any matter fields of spin $\leq 2$. 
We assume the EFT cutoff is parametrically below the Planck scale, $M^2\ll M_{\mathrm{pl}}^2$, meaning that gravity is weakly interacting below $M^2$. 

The amplitudes for various combinations of the two helicity states, with possible permutations, are
\begin{equation}
\begin{aligned}
& \mathcal{M}\left(1^{+} 2^{-} 3^{-} 4^{+}\right)=\langle 23\rangle^4[14]^4 f(s, u) \\
& \mathcal{M}\left(1^{+} 2^{+} 3^{+} 4^{-}\right)=([12][13]\langle 14\rangle)^4 g(s, u) \\
& \mathcal{M}\left(1^{+} 2^{+} 3^{+} 4^{+}\right)=\frac{[12]^2[34]^2}{\langle 12\rangle^2\langle 34\rangle^2} h(s, u)
\end{aligned}
\end{equation}
where we use spinor-helicity variables; see e.g. \cite{Elvang:2013cua}.
The functions $g(s, u)$ and $h(s, u)$ are fully crossing symmetric in $s,t,$ and $u$; however, $f(s, u)$ is only crossing symmetric in $s\leftrightarrow u$. 
The low-energy expansions for these functions are
\begin{equation}
\begin{aligned}
f_{\text {low }}(s, u)= & \frac{8 \pi G}{s t u}+\frac{2 \pi G s u}{t}\left|\widehat{g}_3\right|^2+g_4+g_5 t+g_6 t^2-g_6^{\prime} s u+\ldots \\
g_{\text {low }}(s, u)= & \frac{4 \pi G}{s t u} \widehat{g}_3+\frac{1}{2} \widehat{g}_6^{\prime \prime}+\ldots \\
h_{\text {low }}(s, u)= & 40 \pi G \widehat{g}_3 s t u+\frac{1}{2} \widehat{g}_4\left(s^2+t^2+u^2\right)^2+2 \widehat{g}_5 s t u\left(s^2+t^2+u^2\right) \\
& +\widehat{g}_6\left(s^2+t^2+u^2\right)^3+\widehat{g}_6^{\prime} s^2 t^2 u^2+\ldots
\end{aligned}
\end{equation}
where we assume weak coupling, allowing us to ignore loop contributions.

\subsubsection{Crossing symmetric MHV amplitudes}
To apply the crossing symmetric dispersion relations to the MHV sector, we construct fully crossing symmetric combinations of $f(s,u)$.
These combinations are defined in \cite{Zahed:2021fkp, Mahoux:1974ej,Roskies:1970uj} and were used for pion scattering in \cite{Li:2023qzs},
\begin{equation}
\label{eq:MHVcombinations}
\begin{aligned}
\mathcal{M}^{(1)} & = \bar{f}(s, t) + \bar{f}(t, u) + \bar{f}(s, u) \\
\mathcal{M}^{(2)} & = \frac{\bar{f}(s, t) - \bar{f}(s, u)}{t-u} + \text{cyc perm} \\
\mathcal{M}^{(3)} & = \left( \frac{\bar{f}(s, t) - \bar{f}(s, u)}{t-u} - \frac{\bar{f}(s, t) - \bar{f}(t, u)}{s-u} \right) \frac{1}{s-t} + \text{cyc perm} \,.
\end{aligned}
\end{equation}
We define 
\begin{equation}
\bar{f}(s, t)= u^4 {f}(s, t), \qquad \bar{f}(t, u)= s^4 {f}(t, u), \qquad \bar{f}(s, u)= t^4 {f}(s, u)\,.
\end{equation}
The factors in $\bar f$ are chosen to ensure that the different crossed terms have uniform Regge behavior since, as shown in \cite{Caron_Huot_20232},
\begin{equation}
\lim _{|s| \rightarrow \infty} f\left(s,-p^2\right) \leq C s^{-3}, \quad \lim _{|s| \rightarrow \infty} f\left(s, p^2-s\right) \leq C s 
\end{equation}
for fixed-$u$ and fixed-$t$ respectively, and for some constant $C$. 
The Regge behavior of each amplitude in (\ref{eq:MHVcombinations}) is
\begin{equation}
    \mathcal{M}^{(n)}\leq C s^{k_0^{(n)}}
\end{equation}
where
\begin{equation}
k_0^{(1)}=1, \quad k_0^{(2)}=0, \quad k_0^{(3)}=-1\,.
\end{equation}
The corresponding crossing symmetric sum rules are
\begin{equation}
\label{eq:cssumrulesMHV}
\begin{split}
B_k^{(1)} &= \oint_{z=z_1,z_2,z_3} \frac{d z}{4 \pi i} \mathcal{K}_{k}(z) \mathcal{M}^{(1)}\left(z, p^2\right) \equiv 0 \quad \text{($k\geq 2$ even),} \\
B_k^{(2)} &= \oint_{z=z_1,z_2,z_3} \frac{d z}{4 \pi i} \mathcal{K}_{k-1}(z) \mathcal{M}^{(2)}\left(z, p^2\right) \equiv 0  \quad \text{($k \geq 3$ odd),} \\
B_k^{(3)} &= \oint_{z=z_1,z_2,z_3} \frac{d z}{4 \pi i} \mathcal{K}_{k-2}(z) \mathcal{M}^{(3)}\left(z, p^2\right) \equiv 0  \quad \text{($k\geq 2$ even)}\,.
\end{split}
\end{equation}
For the remaining fully crossing symmetric helicity amplitudes, the high-energy behavior at fixed-$u$ is
\begin{equation}
\lim _{|s| \rightarrow \infty} g\left(s,-p^2\right) \leq C s^{-3}, \quad \lim _{|s| \rightarrow \infty} h\left(s,-p^2\right) \leq C s\,,
\end{equation}
with respective sum rules
\begin{equation}
B_k^{(4)}=\oint_{z=z_1,z_2,z_3} \frac{d z}{4 \pi i} \mathcal{K}_{k-4}(z) g(z, p^2) \equiv 0 , 
\end{equation}
\begin{equation}
B_k^{(5)}=\oint_{z=z_1,z_2,z_3} \frac{d z}{4 \pi i} \mathcal{K}_{k}(z) h(z, p^2) \equiv 0 
\end{equation}
where $k\geq 2$ even.

\subsubsection{High energy partial waves}
For the high-energy side, we use the partial wave expansion for spinning particles
\begin{equation}
\mathcal{M}\left(1^{h_1} 2^{h_2} 3^{h_3} 4^{h_4}\right)=16 \pi \sum_J(2 J+1) a_J^{\{h\}}(s) d_{h_{12}, h_{34}}^J\left(-\frac{\sqrt{m^2-3 p^2}}{\sqrt{m^2+p^2}}\right) 
\end{equation}
where $h_{i j}=h_i-h_j$ and $h_i$ is the helicity of the corresponding external graviton. The $d_{\alpha, \beta}^J(x)$ are Wigner-d functions:
\begin{equation}
d_{h, h^{\prime}}^J(x)=\mathcal{N}_{h, h^{\prime}}^J\left(\frac{1+x}{2}\right)^{\frac{h+h^{\prime}}{2}}\left(\frac{1-x}{2}\right)^{\frac{h-h^{\prime}}{2}}{ }_2 F_1\left(h-J, J+h+1 ; h-h^{\prime}+1 ; \frac{1-x}{2}\right)
\end{equation}
with normalization \cite{Correia:2020xtr}
\begin{equation}
\mathcal{N}_{h, h^{\prime}}^J=\frac{1}{\Gamma\left(h-h^{\prime}+1\right)} \sqrt{\frac{\Gamma(J+h+1) \Gamma\left(J-h^{\prime}+1\right)}{\Gamma(J-h+1) \Gamma\left(J+h^{\prime}+1\right)}} \,.
\end{equation}
We will use the helicity stripped Wigner-d functions:
\begin{equation}
\tilde{d}_{h^{\prime}, h}^J(x)=\mathcal{N}_{h, h^{\prime}}^J { }_2F_1\left(h-J, J+h+1 ; h-h^{\prime}+1 ; \frac{1-x}{2}\right)\,.
\end{equation}
Unitarity of the S-matrix implies that, after summing over intermediate states $X$, the partial-wave coefficients $a_J^{\{h\}}$ obey the positive semidefinite relation
\begin{equation}
i\left[\left(a_J^{-h_4,-h_3,-h_2,-h_1}(s)\right)^*-a_J^{h_1, h_2, h_3, h_4}(s)\right]=\sum_X\left(a_J^{-h_3,-h_4 \rightarrow X}(s)\right)^* a_J^{h_1, h_2 \rightarrow X}(s)\,.
\end{equation}
Omitting the sum and using the following notation for the imaginary part of the amplitude as done in \cite{Caron_Huot_20232}, we have for the MHV function
\begin{equation}
\operatorname{Im} a_J^{+-+-}(s)=\left|c_{J, s}^{+-}\right|^2, \quad \operatorname{Im} a_J^{++--}(s)=\left|c_{J, s}^{++}\right|^2
\end{equation}
and for the other helicity combinations we have
\begin{equation}
\widetilde{\operatorname{Im}} a_J^{+++-}(s)=c_{J, s}^{++} c_{J, s}^{+-}, \quad \widetilde{\operatorname{Im}} a_J^{++++}(s)=\left(c_{J, s}^{++}\right)^2 .
\end{equation}
Here $\widetilde{\operatorname{Im}}$ denotes the discontinuity across the real axis $\widetilde{\operatorname{Im}}  a \equiv[a(s+i 0)-a(s-i 0)] /(2 i)$.
The full set of partial wave decompositions for the various helicity configurations are then
\begin{equation}
\begin{aligned}
& s=m^2>0: \quad \quad \operatorname{Im} f\left(m^2,-p^2\right)=\frac{16 \pi}{m^8} \sum_{J \geq 4}(2 J+1)\left|c_{J, m^2}^{+-}\right|^2 \tilde{d}_{4,4}^J\left(x\right) \\
& t=m^2>0: \quad \operatorname{Im} f\left(-p^2, p^2-m^2\right)=\frac{16 \pi}{m^8} \sum_{\substack{J \geq 0 \\
\text { even }}}(2 J+1)\left|c_{J, m^2}^{++}\right|^2 \tilde{d}_{0,0}^J\left(x\right)
\end{aligned}
\end{equation}
and for fixed-$u$,
\begin{equation}
\begin{aligned}
& \left.\widetilde{\operatorname{Im}}\, g\left(m^2,-p^2\right)\right|_{s=m^2}=\frac{16 \pi}{m^{12}} \sum_{\substack{J \geq 4 \\
\text { even }}}(2 J+1) c_{J, m^2}^{++} c_{J, m^2}^{+-} \tilde{d}_{4,0}^J\left(x\right) \\
& \left.\widetilde{\operatorname{Im}} \,h\left(m^2,-p^2\right)\right|_{s=m^2}=16 \pi \sum_{\substack{J \geq 0 \\
\text { even }}}(2 J+1)\left(c_{J, m^2}^{++}\right)^2 \tilde{d}_{0,0}^J\left(x\right)\,.
\end{aligned}
\end{equation}
As in sec.~\ref{sec:crosssetup}, the high-energy side of the crossing symmetric sum rules is obtained by deforming the contour onto the physical cuts and inserting the appropriate partial-wave decomposition.
For graviton scattering, we define the heavy averages as
\begin{equation}
\langle(\cdots)\rangle \equiv 16 \sum_J(2 J+1) \int_{M^2}^{\infty} \frac{d m^2}{m^2}(\cdots)\,.
\end{equation}
Explicitly, the first few sum rules are
\begin{equation}
\label{eq:gravitonsumrules}
\begin{aligned}
    &B_2^{(1)}:  &\frac{2(8 \pi G)}{p^2}&=
    \left\langle
    \left(\frac{2m^2+3p^2}{m^6}\right)\left( \bar{f}(s, u)+ \bar{f}(s, t) + \bar{f}(t, u)\right)\right\rangle\\
    &B_2^{(3)}:  &\frac{3(8 \pi G)}{p^2}&=
    \left\langle \left(\frac{2 m^2 + 3 p^2}{m^2 + p^2}\right)\left(\left( \frac{\bar{f}(s, t) - \bar{f}(s, u)}{t-u} - \frac{\bar{f}(s, t) - \bar{f}(t, u)}{s-u} \right) \frac{1}{s-t} + \text{cyc perm}\right)\right\rangle\\
    &B_3^{(2)}:  & -3p^2 g_4&= 
    \left\langle
    \left(\frac{2m^2+3p^2}{m^6}\right) \left(
    \frac{\bar{f}(s, t) - \bar{f}(s, u)}{t-u} + \text{cyc perm} \right)\right\rangle\,.
    \end{aligned}
\end{equation}
For the numerical implementation, we also impose positivity in the large-spin regime. 
At large $J$ and fixed impact parameter $b=2J/m$, the helicity-stripped Wigner-d functions reduce to Bessel functions of the first kind \cite{Caron_Huot_20232},
\begin{equation}
\lim _{J, m \rightarrow \infty} \tilde{d}_{h, h^{\prime}}^J\left(1-\frac{2 p^2}{m^2}\right)=\frac{J_{h-h^{\prime}}(b p)}{(p / m)^{h-h^{\prime}}}\,.
\end{equation}
We use this large-spin limit to impose fixed-impact-parameter constraints in the graviton scattering problem, in direct analogy with the scalar case discussed in sec.~\ref{sec:setup}.

\subsection{Bounds}
\label{sec:GGGGbounds}
We now use the crossing symmetric graviton sum rules in \eqref{eq:gravitonsumrules} to compute bounds on the EFT couplings appearing in the MHV amplitude.
As in the scalar case, we consider two choices of smearing range: $p_{\max}=M$, using the spectral assumption with a mass cutoff at $J_*=60$, and $p_{\max}=M/\sqrt{3}$, for which no additional mass cutoff at $J_*$ is required.
We compare the bounds obtained from the crossing symmetric graviton sum rules with the improved-sum-rule bounds of \cite{Caron_Huot_20232}. 
For this comparison, we use all three crossing symmetric MHV sum rules $B_k^{(1)}$, $B_k^{(2)}$, and $B_k^{(3)}$ in (\ref{eq:cssumrulesMHV}) with
\begin{equation}
   \text{CS sum rules:} \qquad  k\in\{2,3,4,5,6\},\qquad n_{\min}=1,\quad n_{\max}=20\,.
\end{equation}
For the improved-sum-rules, we use the following smearing ranges as chosen in \cite{Caron_Huot_20232}:
\begin{equation}
\begin{array}{c|c|c|c}
\text{Improved sum rules} & k & n_{\min} & n_{\max} \\
\hline
B_k^{(1)} & 2 & 1 & 7 \\
B_k^{(2)} & 2 & 10 & 16 \\
B_k^{(1)} & 3,4,5,6 & 0 & 6 \\
B_k^{(2)} & 4,6 & 0 & 6
\end{array}
\end{equation}
We also include bounds with light matter contributions, by which we mean matter states with masses below the cutoff, $m_\ell<M$, to match the setup used for the improved-sum-rule bounds.
For the amplitude of the light matter fields, we adopt the choice made in \cite{Caron_Huot_20232} to ensure that the same shift in the $g_4$ and $g_5$ coefficients is used; this choice ensures Regge growth at fixed-$s$ to be at spin-4 and at fixed-$t$ at spin-2.  Namely, we take 
\begin{equation}
f_{\text {matter }}(s, u)=\sum_{m_{\ell}<M} \frac{\left|g_0^{++}\left(m_{\ell}\right)\right|^2}{m_{\ell}^2-t}+\sum_{m_{\ell}<M} \frac{\left|g_2^{++}\left(m_{\ell}\right)\right|^2\left(1-\frac{6 s u}{ m_{\ell}^4}\right)}{m_{\ell}^2-t}
\end{equation}
where $m_\ell$ is the mass of the spin-0 and spin-2 contributions and for $s, t,$ and $u$ we use (\ref{stucrossvar}). We take crossing symmetric combinations of $f_{\text {matter }}(s, u)$ as done for the MHV amplitude given in (\ref{eq:MHVcombinations}). 

\begin{figure}
    \centering
    \includegraphics[width=0.9\linewidth]{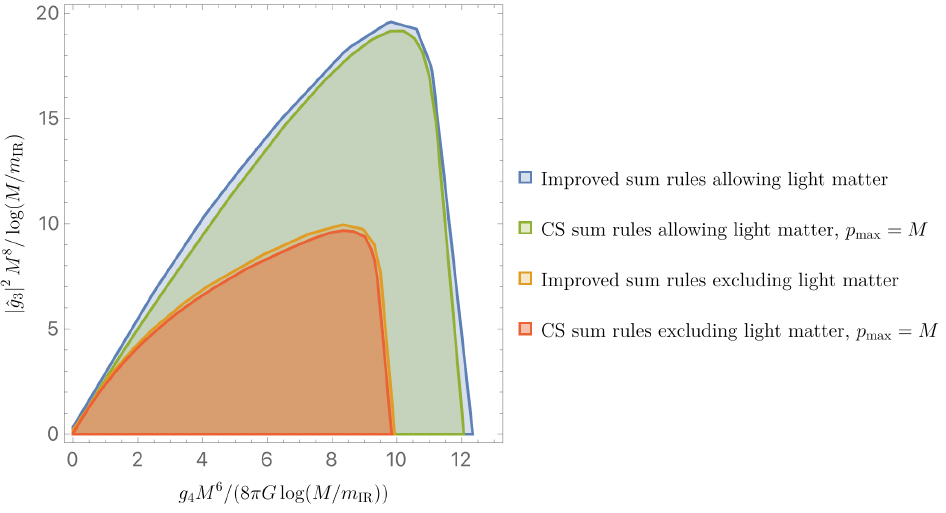}
    \caption{Comparing bounds for $|\hat{g}_3|$ and $g_4$ computed using both crossing symmetric sum rules and improved sum rules, considering both allowing and excluding light matter.}
    \label{fig:MHVcouplingsplotg3g4}
\end{figure}

\begin{figure}
    \centering
    \includegraphics[width=0.9\linewidth]{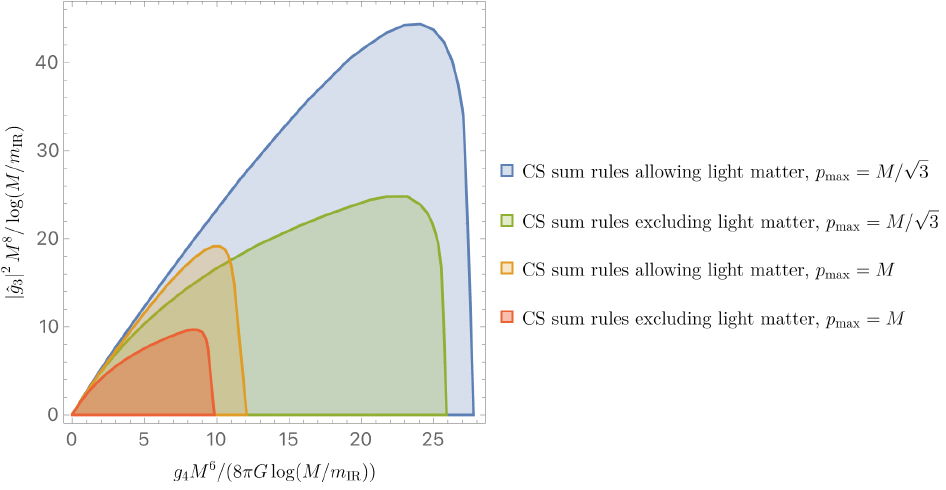}
    \caption{Bounds on $|\hat{g}_3|$ and $g_4$ computed using crossing symmetric sum rules for different choices of $p_\text{max}$ for the momentum smearing. Again, we compute bounds allowing and excluding light matter.}
    \label{fig:MHVcouplingsplotg3g4pmax}
\end{figure}

With the heavy-sector assumptions and light matter contributions specified, we now present the resulting bounds.
Fig.~\ref{fig:MHVcouplingsplotg3g4} shows the results for the $\widehat{g}_3$ and $g_4$ bounds. 
For $p_{\max}=M$, the crossing symmetric sum rules reproduce the bounds obtained from improved sum rules in \cite{Caron_Huot_20232}, providing a nontrivial check of the crossing symmetric construction.
This construction gives a more direct implementation of the gravitational bootstrap bounds, since the required null constraints arise naturally from the crossing symmetric sum rules rather than from combinations of forward-limit sum rules.
As a comparison, we also compute bounds using  $p_\text{max}=M/\sqrt{3}$ shown in fig.~\ref{fig:MHVcouplingsplotg3g4pmax}. 
The smaller smearing range gives weaker bounds with the same overall shape.
For both choices of $p_{\max}$, including light matter enlarges the allowed region relative to the bounds obtained for excluding light matter.

\begin{figure}
\centering
\includegraphics[width=0.8\linewidth]{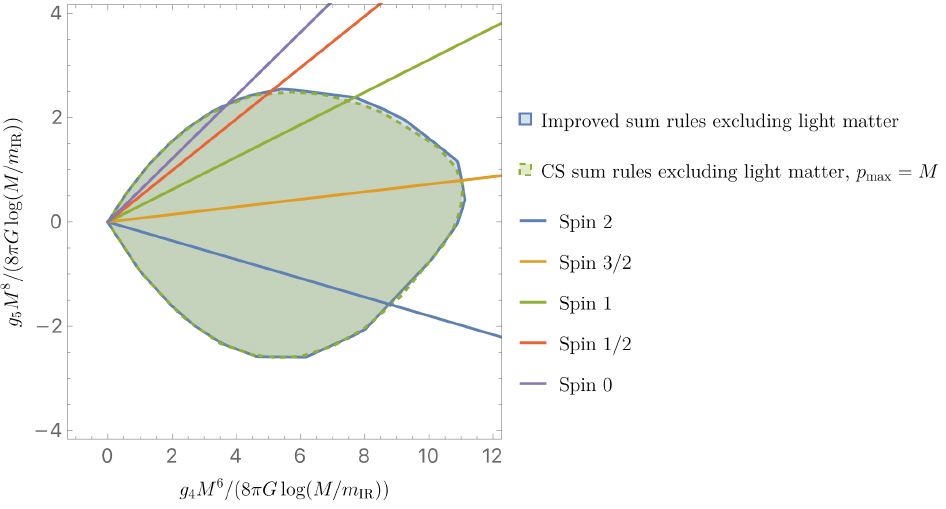}
    \caption{
Allowed region in the coupling parameter space for $g_5$ and $g_4$, comparing crossing symmetric and improved-sum-rule bounds, together with the values generated by one-loop matter contributions from massive particles of spin up to $2$.}
    \label{fig:g5g4plot}
\end{figure}

\begin{figure}
    \centering
    \includegraphics[width=0.8\linewidth]{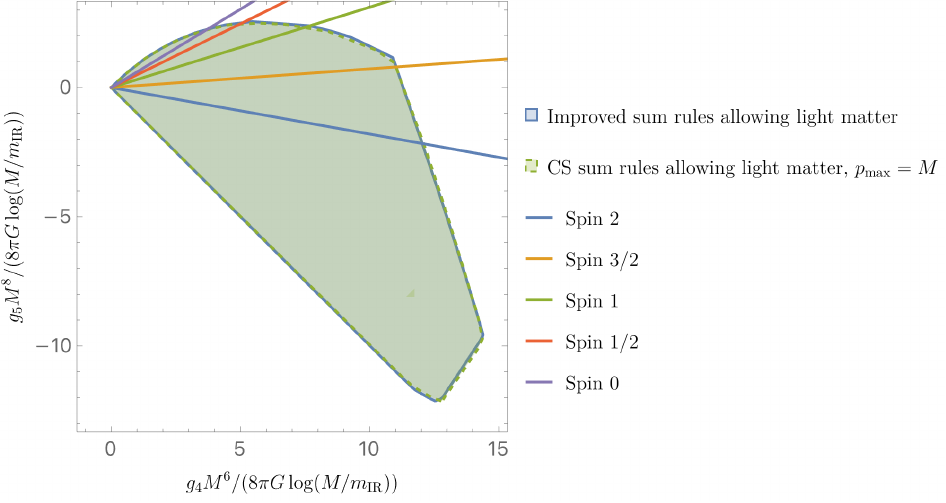}
    \caption{Allowed region for the $g_4$ and $g_5$ couplings allowing light matter fields, computed using both crossing symmetric sum rules with $p_{\max}=M$ and improved sum rules.}
    \label{fig:g5g4CSpmaxwkk}
\end{figure}

\begin{figure}
    \centering
    \includegraphics[width=0.9\linewidth]{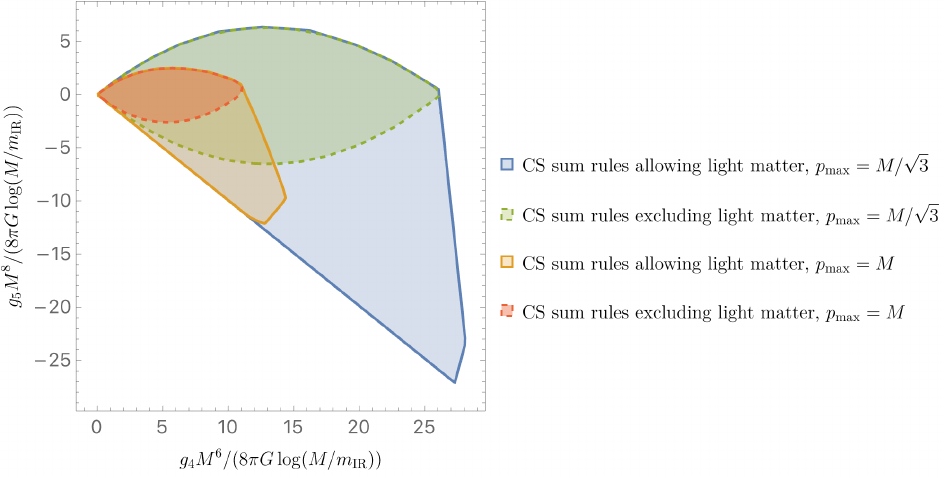}
    \caption{Bounds on the $g_4$ and $g_5$ couplings computed using crossing symmetric sum rules with different momentum-smearing ranges, both allowing and excluding light matter fields.}
    \label{fig:g5g4CScomp}
\end{figure}

Next, we consider bounds on $g_4$ and $g_5$, shown in fig.~\ref{fig:g5g4plot}. 
We include a comparison to explicit models from \cite{Bern:2021ppb}, where the relevant EFT couplings are generated by integrating out massive particles up to spin 2 at one loop. 
The masses of the lightest intermediate states appear at the two-particle threshold, $M=2m$. 
The values of the higher-derivative couplings, taken from \cite{Bern:2021ppb, Caron_Huot_20232}, are:
\[
\begin{array}{lccccc}
\text{Spin} & 0 & \tfrac{1}{2} & 1 & \tfrac{3}{2} & 2 \\
\hline
g_4 M^2 /(G^2 N) & \tfrac{16}{1575} & \tfrac{58}{1575} & \tfrac{248}{1575} & \tfrac{1676}{1575} & \tfrac{5368}{315} \\
g_5 M^4 /(G^2 N) & \tfrac{64}{10395} & \tfrac{944}{51975} & \tfrac{848}{17325} & \tfrac{160}{2079} & -\tfrac{31888}{10395}
\end{array}
\]
where $N$ is the number of particles in the loop. Note that $N$ must be large to be physically significant, since we neglect graviton loops.
The crossing symmetric and improved-sum-rule bounds agree, with the loop examples lying within the resulting allowed region. Fig.~\ref{fig:g5g4CSpmaxwkk} shows the allowed region for $g_4$ and $g_5$ when light matter states are included. 
The crossing symmetric bounds with $p_{\max}=M$ agree with the improved-sum-rule bounds. 
In fig.~\ref{fig:g5g4CScomp}, we compare the crossing symmetric bounds for $p_{\max}=M/\sqrt{3}$ and $p_{\max}=M$, both with and without light matter. 
As in the $\widehat{g}_3$ and $g_4$ bounds, the smaller smearing range gives weaker bounds.

Thus, the crossing symmetric MHV construction developed here provides, to our knowledge, the first implementation of fully crossing symmetric sum rules for amplitudes with spinning external particles.

\subsection{Spin-4 exchange}
\label{sec:spin4exchange}
We now apply the integrated-in massive state setup of sec.~\ref{sec:boundsspectral} to the MHV graviton amplitude in $D=4$. 
Motivated by the massive higher-spin spectrum of string theory, we integrate in a massive spin-4 state below the cutoff and bound its three-point coupling to external gravitons.
After scalar and spin-2 exchanges, this is the first higher-spin state that can appear in the tree-level MHV amplitude.
This spin-4 state is inspired by the massive spectrum from the string Regge trajectory, where for the MHV process the string amplitude is \cite{Bern:2021ppb, Kawai:1985xq}:
\begin{equation}
\mathcal{M}\left(1^{+} 2^{-} 3^{-} 4^{+}\right)=\langle 23\rangle^4[14]^4 \frac{8 \pi G}{s t u} \frac{\Gamma\left(1-\frac{s}{M^2}\right) \Gamma\left(1-\frac{t}{M^2}\right) \Gamma\left(1-\frac{u}{M^2}\right)}{\Gamma\left(1+\frac{s}{M^2}\right) \Gamma\left(1+\frac{t}{M^2}\right) \Gamma\left(1+\frac{u}{M^2}\right)}\left( 1-\frac{s u}{M^2\left(t+M^2\right)} \right)^y
\end{equation}
with closed string tension $\alpha'=4$ and where $y=0,1,2$ corresponds to the superstring, heterotic string and bosonic string respectively.

As in sec.~\ref{sec:sugranewstate}, we raise the cutoff above the mass of the explicitly integrated-in state. 
We denote the mass of the tree-level spin-4 exchange by $m_{J=4}$ and take the cutoff $M$ to correspond to the mass of the next higher-spin state, namely the spin-6 state.
We can evaluate the contribution from the massive spin-4 state on the low-energy side by explicitly including $s, t,$ and $u$-channel poles at $m_{J=4}$. 
For the MHV amplitude, close to the pole, we have:
\begin{equation}
    f(s,u) = \frac{(g_{GG4}^{+-})^2\tilde{d}_{4,4}^4\left(1+\frac{2 u}{s}\right)}{m_{J=4}^2-s}+\frac{(g_{GG4}^{++})^2\tilde{d}_{0,0}^4\left(1+\frac{2 u}{t}\right)}{m_{J=4}^2-t} +\frac{(g_{GG4}^{+-})^2\tilde{d}_{4,4}^4\left(1+\frac{2 s}{u}\right)}{m_{J=4}^2-u} + \text{analytic}\,.
\end{equation}
Equivalently, we can impose delta-functions in the spectral density at mass $m=m_{J=4}$ and spin $J=4$, and evaluate the partial waves using the Wigner-$d$ functions at the chosen mass and spin. 

To compute the bound, we focus on the helicity coupling $(g_{GG4}^{+-})^2$ of the spin-4 state to the external gravitons.
We allow for possible light spin-0 and spin-2 states in the spectrum by including these states as extra positivity conditions in SDPB. These states are allowed to have masses $0<m^2<M^2$. 
Our assumptions for the spectrum below the cutoff are depicted here:
\begin{equation}
\vcenter{\hbox{\scalebox{0.9}{\begin{tikzpicture}
\fill[gray!30] (2.5,0) rectangle (5,6);
    \draw[->] (0,-1) -- (0,6);
    \draw[->] (-1,0)--(5,0);
    \draw (0,2) -- (-0.1,2); 
    \node[anchor=east] at (0,2) {2};
    \draw (0,4) -- (-0.1,4); 
    \fill (0,2) circle[radius=2pt];
    \node[anchor=east] at (0,4) {4};
    \node[anchor=north] at (5,0) {$m^2$}; 
    \node[anchor=east] at (0,6) {$J$};
    \foreach \x in {0,0.25,...,2.5} {
        \fill (\x,2) circle[radius=2pt];
    }
    \foreach \x in {0,0.25,...,2.5} {
        \fill (\x,0) circle[radius=2pt];
    }
    \draw[very thick] (2,0.1) -- (2,-0.2);  
    \node[anchor=north] at (1.9,-0.1) {$m_{J=4}^2$}; 
    \draw[very thick] (2.5,0.1) -- (2.5,-0.2);    
    \node[anchor=north] at (2.7,-0.1) {$M^2$}; 
    \draw[dashed] (2.5,0) -- (2.5,6);
    \fill[blue] (2,4) circle[radius=2pt];
\end{tikzpicture}}}}
\end{equation}
where the blue dot represents the integrated-in spin-4 state and the black dots represent the light spin-0 and spin-2 states. The shaded area represents all states in the heavy averages included in the positivity conditions.

\begin{figure}[h!]
    \centering
    \includegraphics[width=\linewidth]{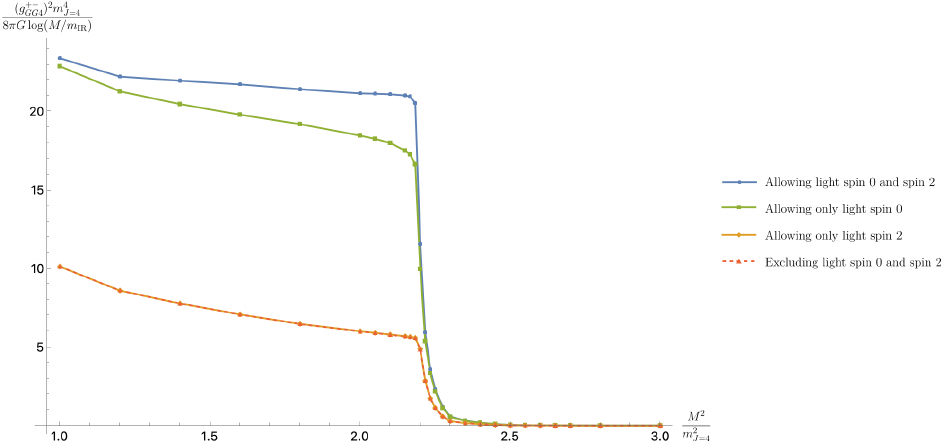}
    \caption{Upper bound on the coupling between external gravitons and a massive spin-4 state including different combinations of light spin-0 and spin-2 states below the cutoff $M$.}
    \label{fig:spin4plot}
\end{figure}

The upper bounds on the coupling $(g_{GG4}^{+-})^2$ of external gravitons to a massive spin-4 state are shown in fig.~\ref{fig:spin4plot} as a function of the ratio $M^2/m_{J=4}^2$. 
A striking feature of the bounds is the sharp drop around $M^2/m_{J=4}^2\simeq 2.2$, where the allowed coupling rapidly approaches zero, regardless of which light low-spin states are included.
A similar sharp feature appeared in the maximal supergravity bound shown in fig.~\ref{fig:sugramphi}.
Near $M^2/m_{J=4}^2=1$, the bounds obtained including light spin-0 states are nearly unchanged by the additional inclusion of light spin-2 states.
As the ratio $M^2/m_{J=4}^2$ increases, the difference between including only light spin-0 states and including both light spin-0 and spin-2 states becomes more pronounced.
Overall, the bound is weaker when light spin-0 states are included.
Including only a massive spin-2 state gives a negligible difference on the bounds compared to excluding both light spin-0 and spin-2 states.
Since the additional light low-spin states are included independently, our spectral assumptions do not impose the correlations among masses and spins that would arise in a string spectrum.
It would be interesting to compare these bounds with those obtained from more string-like spectral assumptions, including leading and daughter Regge trajectories.

\section{Discussion}
\label{sec:discussion}
We derived bounds using crossing symmetric dispersion relations and reproduced known results for scalar scattering with gravity and for graviton scattering in maximal supergravity. 
In the scalar case, smearing the functionals up to $p_\text{max}=M$ rather than $p_\text{max}=M/\sqrt{3}$, as used in \cite{Chang:2025cxc, Beadle:2025cdx}, is essential for exactly reproducing the optimal bounds obtained from improved sum rules.

For $2\rightarrow 2$ graviton scattering in non-supersymmetric theories, we constructed crossing symmetric combinations of the MHV amplitude to study tree-level graviton scattering. 
The construction adapts crossing symmetric combinations used for pion scattering to the graviton case, and ensures the appropriate Regge behavior at each $k$-subtraction.
The crossing symmetric approach offers a more systematic alternative to the improved sum rules used in \cite{sharpboundaries,Caron_Huot_20232}.
We found that the crossing symmetric dispersion relations offer a natural way to write sum rules since they isolate a finite subset of Wilson coefficients, and since by construction they provide null constraints that are usually derived from crossing symmetry.
Using these new sum rules, we reproduced the known graviton-scattering bounds of \cite{Caron_Huot_20232}, providing a fully crossing symmetric implementation of the bootstrap for amplitudes with spinning external states.

We also studied the explicit exchange of massive higher-spin states at tree level by raising the cutoff above the mass of the exchanged state, $M>m_J$, thereby integrating this state into the low-energy amplitude.
We first applied this idea to maximal supergravity, where we reproduced the bounds of \cite{Albert:2024yap} on the coupling of the effective scalar state to external gravitons.
We then used the same approach to study the upper bound for the coupling of a massive spin-4 state exchanged in $2\rightarrow2$ graviton scattering. 
In both the maximal supergravity and spin-4 graviton examples, the bounds exhibit a sharp drop at finite mass ratio, beyond which a nonzero coupling to the integrated-in state is no longer allowed for the assumed spectral setup.
Scanning over the ratio of the spin-4 mass to the cutoff, we found that the bound depends on the assumed light low-spin spectrum, with weaker bounds obtained when light spin-0 states are included.

In this paper, we showed that crossing symmetric dispersion relations provide a useful framework for computing tree-level bounds in gravitational EFTs without relying on the forward limit. 
Although we worked at tree level, the crossing symmetric sum rules are well suited for loop-level calculations where, with or without gravity, the presence of any light particle in EFT loops would give singularities in the forward limit.
Crossing symmetric dispersion relations avoid this issue by working away from the forward limit and therefore provide a natural framework for studying positivity bounds beyond tree level.

For future directions, it would be interesting to study loop effects on the bounds computed here for $2\rightarrow 2$ graviton scattering and test their stability, as was done for scalar scattering in \cite{Beadle:2024hqg,Chang:2025cxc, Beadle:2025cdx}. 
Another direction is to explore whether additional crossing symmetric combinations can be constructed. 
It would also be interesting to study mixed systems, such as graviton-photon scattering, by constructing crossing symmetric combinations of mixed helicity states.

Finally, the construction in \eqref{eq:MHVcombinations} shows that crossing symmetric sum rules can be applied even when the original amplitude is not fully crossing symmetric in $s,t,$ and $u$.
In the MHV case, the amplitude has only a two-variable crossing symmetry, but suitable combinations restore full crossing symmetry at the level of the sum rules.
It would be interesting to apply this idea to amplitudes with even less manifest crossing symmetry.

\section*{Acknowledgements}
We would like to thank Justin Berman, Joydeep Chakravarty, Miguel Correia, Mathieu Giroux, Kelian H\"aring, Yue-Zhou Li, Brian McPeak, Viraj Meruliya, and Julia Pasiecznik for useful discussions.
We especially thank Simon Caron-Huot for countless helpful discussions and invaluable advising. C.P.’s work is supported in parts by the National Science and
Engineering Council of Canada (NSERC), the Canada Research Chair program,
reference number CRC-2022-00421, and the Walter C. Sumner Memorial Fellowship.

\appendix
\section{Numerical implementation}
\label{sec:numerics}
The numerical framework used to compute bounds on Wilson coefficients closely follows \cite{sharpboundaries,Caron_Huot_20232,Albert:2024yap}. We provide more detail here on the various parameter regimes we used.

\begin{figure}[h!]
    \centering
    \includegraphics[width=0.75\linewidth]{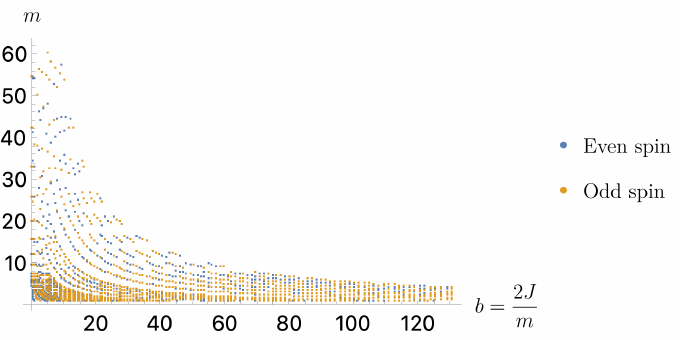}
    \caption{Example of the sampled heavy-state grid in the $(b,m)$ plane used for the finite-mass, finite-spin positivity constraints.}
    \label{fig:heavystategrid}
\end{figure}

\emph{Finite mass, finite spin:} We use a discretization in mass and spin for the spectral density of the high-energy side of the dispersion relations. Positivity must be imposed for each value $m>M$ and all spins $J$ (where only even spins are used for scalar scattering). 
In practice, we include all spins up to $J=60$ and sample more sparsely up to $J_\text{max}=300$.
For the finite-spin constraints, we use a nonuniform mass grid with denser sampling just above the cutoff and sparser sampling at larger masses, implemented using exponentially spaced points. 
For larger spins, the sampling is instead organized more naturally in terms of the impact parameter.
The sampled points are chosen using a nonuniform grid in $(b,m)$, with $b\simeq2J/m$ in the large-spin limit.
Increasing the spin range further or using a denser set of $(m,J)$ states substantially increases the runtime and did not lead to any significant change to the bounds. 
An example of the heavy-state grid used in this regime is shown in fig.~\ref{fig:heavystategrid}.

\emph{Large mass, fixed impact parameter:} 
We also impose constraints in the large-mass limit at fixed impact parameter, following \cite{sharpboundaries}. 
At fixed impact parameter $b$, the high-energy limit of the partial waves is given by the Bessel-function approximation in \eqref{besselfunc}. 
We include the fixed-impact-parameter constraints for $b<b_{\max}$ by discretizing in $b$ and evaluating the heavy averages using the corresponding impact-parameter functionals. 
In the numerics, we take
\begin{equation}
    b_{\max}=200\,.
\end{equation}

\emph{Large mass, large impact parameter:} 
To impose positivity at larger impact parameter, $b>b_{\max}$, we use the large-$b$ form of the smeared Bessel functionals, where for scalar scattering,
\begin{equation}
\Gamma\left(\frac{D-2}{2}\right) \int_0^1 d p \, f(p)
\frac{J_{\frac{D-4}{2}}(b p)}{(b p / 2)^{\frac{D-4}{2}}}
=
A(b)+B(b) \cos \left(b-\frac{\pi(D-1)}{4}\right)
+C(b) \sin \left(b-\frac{\pi(D-1)}{4}\right)\,.
\end{equation}
In practice, we impose positivity by writing
\begin{equation}
A(b)+B(b) \cos \phi+C(b) \sin \phi
=
\left(\cos \frac{\phi}{2} \ \ \sin \frac{\phi}{2}\right)
M(b)
\binom{\cos \frac{\phi}{2}}{\sin \frac{\phi}{2}}\,,
\end{equation}
where
\begin{equation}
M(b)=
\left(\begin{array}{cc}
A(b)+B(b) & C(b) \\
C(b) & A(b)-B(b)
\end{array}\right)\,.
\end{equation}
Thus, for the large-$b$ linear constraints, we impose the positive semidefiniteness of $M(b)$. 
For spinning external states, we use the corresponding modification described in Appendix C of \cite{Caron_Huot_20232}. 

Together, these regimes define the numerical positivity conditions used to compute the coupling bounds.

\newpage

\providecommand{\href}[2]{#2}\begingroup\raggedright\endgroup

\end{document}